\documentclass[letter,bibyear]{aa} % for the letters

\usepackage{epsfig}
\usepackage{amsmath}
\usepackage{subfigure}
\usepackage{natbib}
\usepackage{multicol}
\usepackage{multirow}
\usepackage{color}
\usepackage{float}
\usepackage{longtable}
\usepackage{graphicx}
\usepackage{epstopdf}
\usepackage{booktabs}
\usepackage{txfonts}
\newcommand{\jms}{J.~Mol.~Spectrosc.}   % Journal of Molecular Spectroscopy
\newcommand{\jmst}{J.~Mol.~Struct.}   % Journal of Molecular Structure

\newcommand{\kms}{km s$^{-1}$}

\newcommand{\prev}{Phys. Rev.}
\begin{document}

\title{Discovery of C$_5$H$^+$ and detection of C$_3$H$^+$ in TMC-1 with the 
QUIJOTE$^1$ line survey
\thanks{Based on observations carried out
with the Yebes 40m telescope (projects 19A003,
20A014, 20D023, and 21A011). The 40m
radiotelescope at Yebes Observatory is operated by the Spanish Geographic 
Institute
(IGN, Ministerio de Transportes, Movilidad y Agenda Urbana).}}

\author{
J.~Cernicharo\inst{1},
M.~Ag\'undez\inst{1},
C.~Cabezas\inst{1},
R.~Fuentetaja\inst{1},
B.~Tercero\inst{2,3},
N.~Marcelino\inst{2},
Y.~Endo\inst{4},
J.~R.~Pardo\inst{1}, and
P.~de~Vicente\inst{3}
}

\institute{Grupo de Astrof\'isica Molecular, Instituto de F\'isica Fundamental (IFF-CSIC),
C/ Serrano 121, 28006 Madrid, Spain\\ \email jose.cernicharo@csic.es
\and Centro de Desarrollos Tecnol\'ogicos, Observatorio de Yebes (IGN), 19141 Yebes, Guadalajara, Spain
\and Observatorio Astron\'omico Nacional (OAN, IGN), Madrid, Spain
\and Department of Applied Chemistry, Science Building II, National Chiao Tung University, 1001 Ta-Hsueh Rd., Hsinchu 30010, Taiwan}

\date{Received; accepted}

\abstract{
We report the discovery of the C$_5$H$^+$ cation toward TMC-1 with the QUIJOTE$^1$ line survey.
Four lines from $J$=7-6 up to $J$=10-9 have been identified in perfect harmonic frequency relation
that can be fit with $B$=2411.94397$\pm$0.00055 MHz and $D$=138$\pm$3 Hz. The standard deviation
of the fit is 4.4 kHz. After discarding potential candidates, C$_5$H$^-$ among them, we conclude that 
the carrier is C$_5$H$^+$
, for which accurate ab initio calculations provide $B$=2410.3 MHz. We also report for the first time in a cold starless core the detection of the C$_3$H$^+$ cation. The column densities we derive for
C$_5$H$^+$ and C$_3$H$^+$ are (8.8$\pm$0.5)$\times$10$^{10}$ cm$^{-2}$ and (2.4$\pm$0.2)$\times$10$^{10}$ cm$^{-2}$,
respectively. Hence, the C$_5$H$^+$/C$_3$H$^+$ abundance ratio is 3.7$\pm$0.5. The fact that C$_5$H$^+$ is more
abundant than C$_3$H$^+$ is well explained by dedicated chemical models and is due to 
the slow reactivity of C$_5$H$^+$ with H$_2$, while C$_3$H$^+$ reacts with H$_2$.
}

\titlerunning{C$_5$H$^+$ in TMC-1}
\authorrunning{Cernicharo et al.}

\maketitle

\section{Introduction}

The QUIJOTE\footnote{\textbf{Q}-band \textbf{U}ltrasensitive \textbf{I}nspection \textbf{J}ourney 
to the \textbf{O}bscure \textbf{T}MC-1 \textbf{E}nvironment} 
line survey of TMC-1 \citep{Cernicharo2021a} performed with the Yebes 40m radio telescope has
permitted detection of nearly 27 new molecular species in the last months, most of them hydrocarbons and cycles,
indene among them (see, e.g., \citealt{Agundez2021,Cernicharo2021b,Cernicharo2021c} and references therein). QUIJOTE
has now reached a level of sensitivity that permits performing rotational spectroscopy of unknown species in space. Several molecules have
been detected in this source that lack previous rotational spectroscopic laboratory information. Based on spectral patterns
found in the data that can be assigned to linear or asymmetric species, and with the help of a high 
level of theory quantum chemical calculations, we have discovered molecules such as HC$_5$NH$^+$ \citep{Marcelino2020}, 
HC$_3$O$^+$\citep{Cernicharo2020a}, HC$_3$S$^+$ \citep{Cernicharo2021d}, CH$_3$CO$^+$ \citep{Cernicharo2021e},
HCCS$^+$ \citep{Cabezas2021a}, and HC$_7$NH$^+$ \citep{Cabezas2022}.
%, HCCNCH$^+$ \citep{Agundez2022}. \textcolor{blue}{(I would skip HCCNCH$^+$ until it is submitted)}

TMC-1 and IRC+10216 are the sources in which all known C$_n$H$^-$ anions in space have been detected
\citep{McCarthy2006,Cernicharo2007,Brunken2007,Kawaguchi2007,Remijan2007}. Nitrile anions CN$^-$, C$_3$N$^-$ and
C$_5$N$^-$ were first detected in the circumstellar envelope of IRC+10216 \citep{Agundez2010,Thaddeus2008,Cernicharo2008}.
Only with the sensitivity of QUIJOTE has it been possible to detect the anions C$_3$N$^-$ and C$_5$N$^-$ in TMC-1
\citep{Cernicharo2020b}. The detection of new cations and anions is an astounding source of information with which chemical models can be improved and insights into the chemical paths can be gained that lead to their formation.

In this Letter we report the discovery of four strong lines in TMC-1, with a perfect harmonic relation, that we assign to the 
C$_5$H$^+$ cation. 
The only alternative plausible candidate, C$_5$H$^-$,
has been ruled out on the basis of accurate ab initio calculations and the lack of emission of these lines in IRC+10216.
In addition, we report for the first time the
detection of the C$_3$H$^+$ cation in a starless core.   C$_3$H$^+$ was previously detected
only toward PDRs
%photodissociation regions 
\citep{Pety2012,McGuire2014,Cuadrado2015,Guzman2015}, Sgr\,B2 \citep{McGuire2013}, 
diffuse clouds \citep{Gerin2019}, and the z=0.89 source PKS 1830-211 \citep{Tercero2020}. A previous claim of 
detection of a line of C$_3$H$^+$ in absorption toward \mbox{TMC-1} \citep{McGuire2013} is ruled out by our 
sensitive QUIJOTE observations, which clearly show this line to be in emission. We use dedicated chemical models
to compare the expected abundances with the observed ones and obtain excellent agreement 
between models and observations.

\section{Observations} \label{observations}

New receivers, built within the Nanocosmos project\footnote{\texttt{https://nanocosmos.iff.csic.es/}}
and installed at the Yebes 40m radiotelescope, were used
for the observations of TMC-1
($\alpha_{J2000}=4^{\rm h} 41^{\rm  m} 41.9^{\rm s}$ and $\delta_{J2000}=
+25^\circ 41' 27.0''$). A detailed description of the system is 
given by \citet{Tercero2021}.
The receiver consists of two cold high electron mobility transistor amplifiers covering the
31.0-50.3 GHz band with horizontal and vertical             
polarizations. Receiver temperatures in the runs achieved during 2020 vary from 22 K at 32 GHz
to 42 K at 50 GHz. Some power adaptation in the down-conversion chains reduced
the receiver temperatures during 2021 to 16\,K at 32 GHz and 25\,K at 50 GHz.
The backends are $2\times8\times2.5$ GHz fast-Fourier transform spectrometers
with a spectral resolution of 38.15 kHz
providing the whole coverage of the Q band in both polarizations. 
All observations were performed in frequency-
switching mode with frequency throws of 8 and 10 MHz. The main beam efficiency varies from 0.6 at
32 GHz to 0.43 at 50 GHz. The intensity scale used in this work, antenna temperature
($T_A^*$), was calibrated using two absorbers at different temperatures and the
atmospheric transmission model ATM \citep{Cernicharo1985, Pardo2001}.
Calibration uncertainties were adopted to be 10~\%.
All data were analyzed using the GILDAS package\footnote{\texttt{http://www.iram.fr/IRAMFR/GILDAS}}.
Details of the QUIJOTE line survey are provided by \citet{Cernicharo2021a}. The data presented here
correspond to 238 hours of observing time on the source.

\begin{figure}
\centering
\includegraphics[scale=0.58]{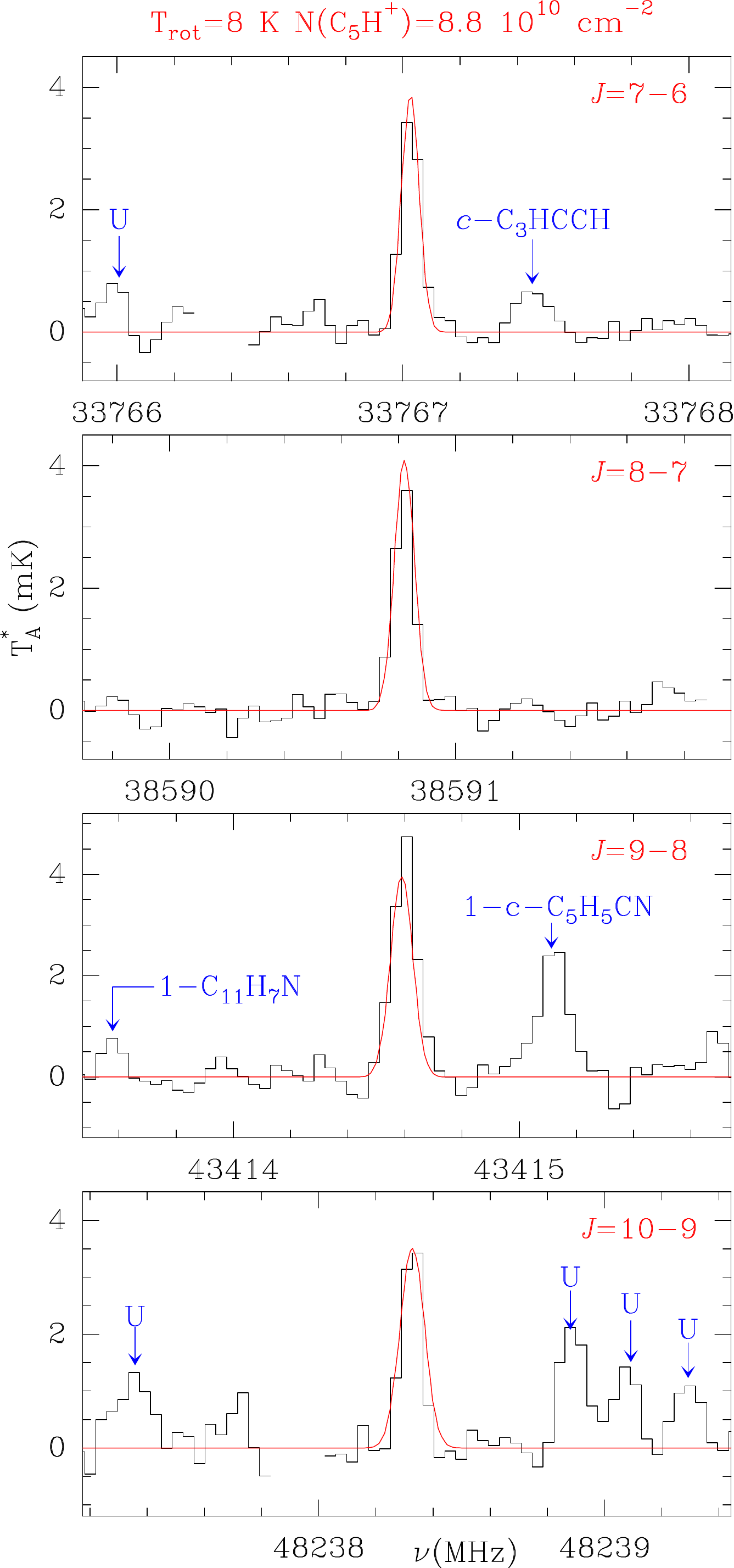}
\caption{Observed lines of C$_5$H$^+$ toward TMC-1.
Line parameters are given in Table \ref{line_parameters}.
The abscissa corresponds to the rest frequency assuming a local standard of rest velocity of 5.83
km s$^{-1}$. 
The ordinate is the antenna temperature corrected for atmospheric and telescope losses in mK.
The red line shows the synthetic spectrum derived for
T$_{rot}$=8\,K and N(C$_5$H$^+$)=8.8$\times$10$^{10}$ cm$^{-2}$.
Blank channels correspond to negative features produced
in the folding of the frequency-switching data.}
\label{fig_c5h+}
\end{figure}

\section{Results} \label{results}

Line identification in this work was done using the catalogs 
MADEX \citep{Cernicharo2012}, CDMS \citep{Muller2005}, and JPL \citep{Pickett1998}. 
By December 2021, the MADEX code contained 6421 spectral
entries corresponding to the ground and vibrationally excited states, together
with the corresponding isotopologs, of 1727 molecules. 

With the current level of sensitivity of QUIJOTE, we have detected 
1591 features above the 1\,mK level (5 $\sigma$ below 42 GHz). Of these lines, 188 remain unidentified. 
We note, however, that the number of unknown spectral features above the 3$\sigma$ level is much larger.
Future improved QUIJOTE data will permit us to confirm them. Only four of these unidentified lines have intensities above 3 mK. The
frequencies of these
four lines are, in addition, in perfect harmonic relation with $J_u$=7, 8, 9, 10. They are
shown in Fig. \ref{fig_c5h+} and their line parameters are given in Table \ref{line_parameters}. They
do not show any hyperfine structure, and no other nearby lines are present with similar intensities. This discards a symmetric rotor, or a linear radical, as possible carrier. This is therefore a linear molecule with a $^1\Sigma$ ground electronic state, or with a slightly
asymmetric rotor with electronic state $^1A$. 
By fitting the observed frequencies to the standard Hamiltonian for a linear molecule 
($\nu(J\rightarrow J-1)=2B_0J-4D_0J^3$), we derived $B_0$=2411.94397$\pm$0.00055 MHz and $D_0$=138$\pm$3 Hz. The standard deviation
of the fit is 4.4 kHz. The possibility that these four strong lines 
appear by chance in harmonic relation with this precision is negligible. This means that we have detected a new molecular species
in TMC-1.

\begin{table*}
\centering
\caption{Observed lines of C$_5$H$^+$ in TMC-1.}
\label{line_parameters}
\begin{tabular}{cccccc}
\hline
Transition&$\nu_{obs}$$^a$    &$\nu_{obs}-\nu_{cal}$$^b$&$\int$ $T_A^*$ dv $^c$  &$\Delta$v\,$^d$& $T_A^*$\,$^e$ \\
\hline              
7 - 6     &33767.026$\pm$0.005& -0.2                    &2.87$\pm$0.18           & 0.74$\pm$0.06& 3.65$\pm$0.16\\
8 - 7     &38590.818$\pm$0.005& -2.8                    &2.60$\pm$0.12           & 0.67$\pm$0.04& 3.66$\pm$0.17\\
9 - 8     &43414.594$\pm$0.005&  5.1                    &3.39$\pm$0.17           & 0.69$\pm$0.04& 4.60$\pm$0.21\\
10 - 9    &48238.325$\pm$0.005& -2.2                    &2.28$\pm$0.24           & 0.54$\pm$0.07& 3.98$\pm$0.41\\
\hline                                                                                             
\end{tabular}
\tablefoot{\\
\tablefoottext{a}{Observed frequency assuming a v$_{LSR}$ of 5.83 \kms.}
\tablefoottext{b}{Observed minus calculated frequencies in kHz.}
\tablefoottext{c}{Integrated line intensity in mK\,km\,s$^{-1}$.}
\tablefoottext{d}{Line width at half-intensity derived by fitting a Gaussian function to
the observed line profile (in km\,s$^{-1}$).}
\tablefoottext{e}{Antenna temperature in millikelvin.}
}
\end{table*}

The rotational constants $B$ and $D$ of the new species are very close to those of C$_5$H,
for which $B$=2395.131$\pm$0.001 MHz and $D$=127.41$\pm$0.03 Hz \citep{Cernicharo1986a,Cernicharo1986b,
Cernicharo1987,Gottlieb1986,McCarthy1999}. The lack of fine or hyperfine structure permits us to
exclude radical molecules. Nevertheless, and in order to explore what type of 
candidates could fit the observed $B$ and $D$, we note that C$_4$N has a rotational constant of 2424.36 MHz, but its ground
electronic state is $^2\Pi_r$ \citep{McCarthy2003}; for C$_4$O, the rotational constant is 2351.26 MHz, but its electronic
ground state is $^3\Sigma$ \citep{Ohshima1995}. Finally, the HC$_4$O species is 
also a radical, with a rotational constant of 2279.91 MHz \citep{Kohguchi1994}. The rotational constant of species
containing a sulphur atom such as HC$_3$S$^+$ is too large, $B$=2735.46 MHz \citep{Cernicharo2021d}.
The same holds for NCCS, which is a bent radical with $(B+C)/2$=2829.36 MHz \citep{Nakajima2003}.
Concerning asymmetric species, HNCCS has been calculated to have an electronic
ground state $^1A$ \citep{Gronowski2015}, and it is an isomer 147 kJ/mol above of HCSCN, a molecule
recently detected in TMC-1 with QUIJOTE \citep{Cernicharo2021f}. However, although it is slightly 
asymmetric ($A$=617.9 GHz), the value calculated by \citet{Gronowski2015} for $(B+C)/2$ is 2563 MHz, which
represents a deviation of 6.3\% with respect to the rotational constant of our species.
The possibility that the lines belong to an isotopolog $^{13}$C, $^{34}$S, or D of an abundant species 
harboring line intensities above 300 mK (corresponding to a $^{13}$C substitution) or 100 mK (for a $^{34}$S or D substitution) is
discarded from the analysis of all abundant species that were previously identified in the QUIJOTE data. 
It therefore appears that the carrier could contains five carbon atoms, or four carbon atoms and one nitrogen or oxygen
atom. The only species that could fit these requirements are C$_5$H$^+$ and C$_5$H$^-$ as all 
the other plausible candidates with O or N are open shell species. It might be argued that the carrier could be
one of the anions of these radicals. However, none of the neutral radical species quoted previously has been detected in TMC-1.
Moreover, molecules with a single sulphur atom do not fit the derived rotational constant.

Quantum chemical calculations by \citet{Botschwina1991} provided a rotational constant for C$_5$H$^+$ of $B_0$=2405$\pm$5 MHz, and
a proton affinity for C$_5$ of 860$\pm$5 kJ/mol. A value of $B_e$ for this species of $\sim$2420 MHz was obtained
by \citet{Aoki2014}. More recent calculations by \citet{Bennedjai2019} provided a value 
for $B_e$ of 2391.7 MHz. 
To obtain accurate spectroscopic parameters, we performed ab initio calculations for C$_5$H$^+$ and also for the C$_5$H species, 
for which the rotational parameters were experimentally determined. In this manner, we can scale the calculated values for the 
C$_5$H$^+$ species using experimental/theoretical ratios derived for the related species C$_5$H. This procedure has been found to provide 
rotational constants with an accuracy better than 0.1\,\% (e.g., \citealt{Cabezas2021b}). The geometry optimization calculations for all 
the species were made using the coupled cluster method with single, double, and perturbative triple excitations with an explicitly correlated 
approximation (CCSD(T)-F12; \citealt{Knizia2009}) and all electrons (valence and core) correlated together with the Dunning correlation 
consistent basis sets with polarized core-valence correlation triple-$\zeta$ for explicitly correlated calculations (cc-pCVTZ; \citealt{Hill2010}). 
These calculations were carried out using the Molpro 2020.2 program \citep{Werner2021}. The values for the centrifugal distortion constants were 
obtained using harmonic vibrational frequency calculations, with the second-order M{\o}ller-Plesset perturbation theory method 
(MP2; \citealt{Moller1934}) and the correlation consistent with polarized valence triple-$\zeta$ basis set (cc-pVTZ; \citealt{Woon1993}). 
These calculations were carried out using the Gaussian09 program \citep{Frisch2016}. 
The derived results are presented in Table \ref{abini}.

The comparison of the calculated and experimental rotational constant for C$_5$H species reveals the good accuracy of the ab initio 
calculations we employed. The differences between the calculated $B_e$ value and the experimental value is about 0.2\% for C$_5$H. The scaled 
value for C$_5$H$^+$ is 2410.3\,MHz, which matches the observed rotational constant very well, the difference is 0.07\%. The scaled value 
for the distortion constant C$_5$H$^+$ also reproduces the $D$ value derived from the \mbox{TMC-1} lines remarkably well. 
Hence, our ab initio calculations provide 
conclusive arguments for the spectroscopic identification of C$_5$H$^+$ using our sensitive QUIJOTE survey
of \mbox{TMC-1}. Using the $B_e$ value
calculated by \citet{Bennedjai2019} for C$_5$H, and correcting $B_e$ of C$_5$H$^+$ by their ratio $B_{obs}(C_5H)/B_e(C_5H)$, we found 
$B_0(C_5H^+)\sim$ 2408.3 MHz, which also agrees well with our experimental rotational constant.

\begin{table*}
\small
\caption{Theoretical spectroscopic parameters for the different molecular candidates for the observed lines in TMC-1 (all in MHz).}
\label{abini}
\centering
\begin{tabular}{{lccccccccc}}
\hline
\hline
& &\multicolumn{2}{c}{C$_5$H$^+$ ($^1\Sigma$)} &\multicolumn{2}{c}{$l$-C$_5$H$^-$ ($^3\Sigma$)} &\multicolumn{2}{c}{$ql$-C$_5$H$^-$ ($^1A$)}&\multicolumn{2}{c}{C$_5$H ($^2\Pi$)}\\
\cmidrule(lr){3-4} \cmidrule(lr){5-6} \cmidrule(lr){7-8} \cmidrule(lr){9-10}
Parameter & TMC-1\tablefootmark{a} & Calc.\tablefootmark{b} & Scaled\tablefootmark{c} &  Calc.\tablefootmark{b} & Scaled\tablefootmark{c} & Calc.\tablefootmark{b} & Scaled\tablefootmark{c} & Exp.\tablefootmark{d} & Calc.\tablefootmark{b} \\
\hline
$B_e$              &  2411.94397(55)& 2404.2 &  2410.3 &  2366.4   &  2372.4  &   2389.5  &  2395.5\tablefootmark{e}  &    2395.131(1) &  2389.1 \\
$D$ x 10$^{-6}$    &  138(3)        & 102    &  127    &    97.9   &  121     &   101     &  126     &     127.41(3)  &   103     \\
\hline
\end{tabular}
\tablefoot{\tablefoottext{a}{Parameters derived using the TMC-1 frequencies
from Table \ref{line_parameters}.} \tablefoottext{b}{This work; see text.} \tablefoottext{c}{This work; scaled by the ratio Exp/Calc. of the corresponding parameter for C$_5$H species}. 
\tablefoottext{d}{\citet{Cernicharo1986a,Cernicharo1986b,Cernicharo1987,Gottlieb1986,McCarthy1999}. \tablefoottext{e}{($B+C$)/2.}} 
}\\
\end{table*}

\begin{figure}
\centering
\includegraphics[scale=0.72]{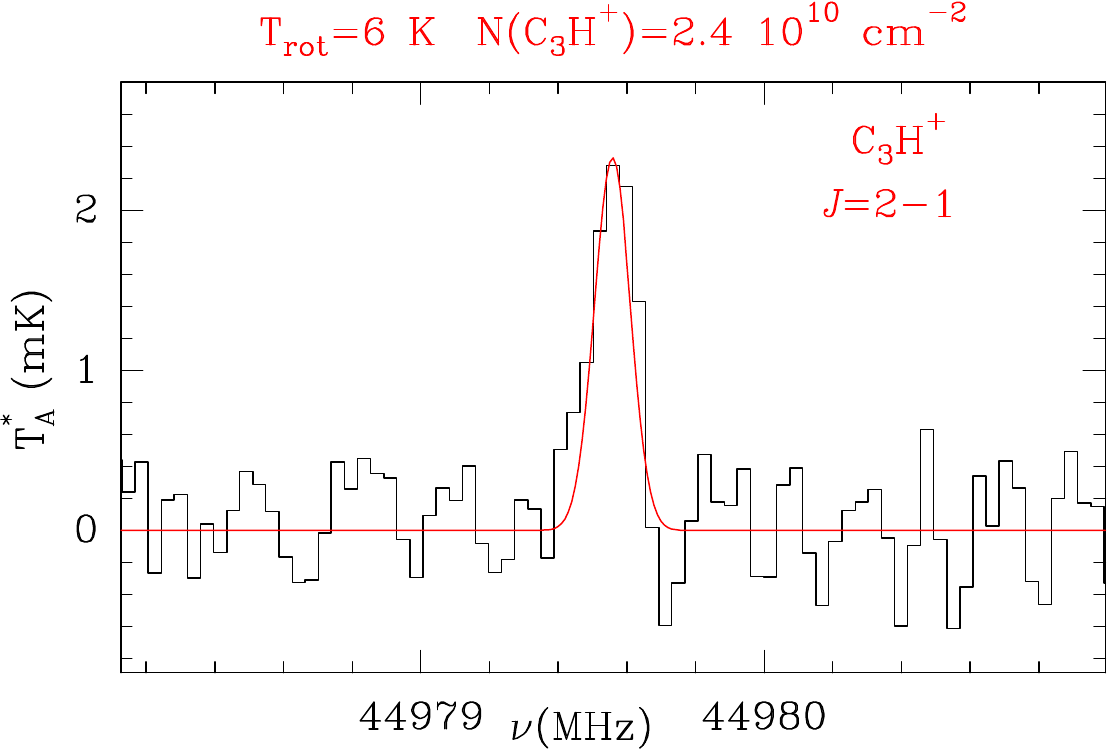}
\caption{Observed $J$=2-1 line of C$_3$H$^+$ toward TMC-1.
The abscissa corresponds to the rest frequency assuming a local standard of rest velocity of 5.83
km s$^{-1}$. 
The ordinate is the antenna temperature corrected for atmospheric and telescope losses in mK.
The red line shows the synthetic spectrum derived for
T$_{rot}$=6\,K and N(C$_3$H$^+$)=2.4$\times$10$^{10}$ cm$^{-2}$.
}
\label{fig_c3h+}
\end{figure}

The other species that might fit the observed rotational constant is C$_5$H$^-$. However, there is a significant controversy
about the nature of the ground state of this species. Two-color photodetachment studies by \cite{Tulej2011} suggested that
the ground state is $^3\Sigma^-$ with a linear structure and a rotational constant of 2476.3$\pm$6 MHz. 
In addition, the reported experimental 
value for the C$_5$H$^-$ ($^3\Sigma^-$ ) species differs by more than 100 MHz from the predicted values obtained by 
\citet{Bennedjai2019} and our scaled value (Table \ref{abini}). Moreover, if that assignment were correct, 
C$_5$H$^-$ could not be the carrier of the observed lines as significant hyperfine structure is expected due 
to the spin 1/2 of H and the low  spin-spin interaction constant derived, $\lambda$=7200$\pm$4500 MHz. 
Consequently, the transitions with $J_u=N_u+1$ and $J_u=N_u-1$ will be close to those corresponding to $J_u=N_u$, and
the $J_u=N_u+1$ lines should be stronger than those with $J_u=N_u$. No lines with such a pattern
are observed in our survey. Hence, we discard this putative state of C$_5$H$^-$ as the carrier of our lines. There is also an
additional caveat for this triplet linear C$_5$H$^-$. All linear anions detected so far in space have $B$(anion)<$B$(neutral). 
However, for the molecule observed by \cite{Tulej2011}, the situation is the opposite, $B$(anion)>$B$(neutral). 
Additional controversy arises from the recent calculations of \citet{Bennedjai2019}. They concluded that the ground
electronic state of C$_5$H$^-$ contains a C$_3$ ring with very different rotational constants than those observed. 
In addition, they also provided strong arguments against the viability of the triplet linear form as it is
very dependent on the electron correlation energy, which denotes instability for this state. Hence, 
the ring ground electronic state of C$_5$H$^-$, although potentially expected in TMC-1, is not the carrier of our lines. 
We note, however, that they also computed a second slightly asymmetric isomer for this anion.
It is 0.7 eV above the ring structure, with rotational constants $A_0\sim$793 GHz, and $(B+C)/2\sim$ 2395.2 MHz. This
bent $^1A$ isomer could also fit the observed rotational constant. Our calculations predict 
a scaled value for $(B+C)/2\sim$2395.5 MHz (see Table \ref{abini}).

All the anions that were detected in TMC-1 were also observed in the envelope of the carbon-rich star IRC+10216.
We have explored the recent survey of this source performed with the Yebes 40m radio telescope \citep{Pardo2021} and
found no emission at the frequencies of the four lines of TMC-1. We also searched
in the available data from the IRAM 30m telescope at 3mm and 2mm for lines in these ranges without success.
Although this is not a definitive argument, it is an important drawback for bent-C$_5$H$^-$ ($^1A$)
being the carrier of the lines. 
Finally, although C$_3$H$^+$ has been found only 
in PDRs and diffuse media so far \citep{Pety2012,McGuire2013,McGuire2014,Cuadrado2015,Guzman2015,Gerin2019,Tercero2020}, its presence in TMC-1 could be a solid
argument to assign C$_5$H$^+$ as carrier of the U-lines. Using the QUIJOTE data, a very nicely detected line (S/N$\sim$10) 
appears just at the predicted frequency of the $J=$2-1 transition of C$_3$H$^+$. The line is shown in Fig. \ref{fig_c3h+}. 
This represents the first
detection of this cation in a cold starless core and provides key information about the chemistry of C$_3$H$^+$ in
cold and dense environments. Moreover, the observation of the $J=$2-1 line of C$_3$H$^+$ in emission rules out 
a previous tentative identification of this species in \mbox{TMC-1} based on a noisy absorption feature at the frequency 
of this same transition \citep{McGuire2013}.

Adopting C$_5$H$^+$ as carrier of the U-lines, a model line profile fitting procedure provides a rotational temperature
of 8$\pm$1\,K and a column density of (8.8$\pm$0.5)$\times$10$^{10}$ cm$^{-2}$. 
We assumed a dipole moment value of 2.88\,D \citep{Botschwina1991}.
The computed synthetic line spectra are
shown in Fig. \ref{fig_c5h+} and fit the observed line intensities and profiles very well. We adopted
a line width of 0.6\,\kms\  and a uniform brightness source of diameter 80$''$ \citep{Fosse2001}. For C$_3$H$^+$
we have only one line, and we adopted a rotational temperature of 6\,K, which provides a column density
for this species of (2.4$\pm$0.2)$\times$10$^{10}$ cm$^{-2}$.
When we assume a column density for H$_2$ of 10$^{22}$ cm$^{-2}$ \citep{CernicharoGuelin}, the abundances
of C$_5$H$^+$ and C$_3$H$^+$ are (8.8$\pm$0.5)$\times$10$^{-12}$ and (2.4$\pm$0.2)$\times$10$^{-12}$, respectively.
The abundance ratio of the two species is 3.5$\pm$0.5.

\section{Discussion}

% proton affinity of C3 and C5, general behavior of protonated molecules and trend of increasing protonated-to-neutral ratio vs proton affinity (refer to Agundez et al 2015).

\begin{figure}
\centering
\includegraphics[width=0.46\textwidth]{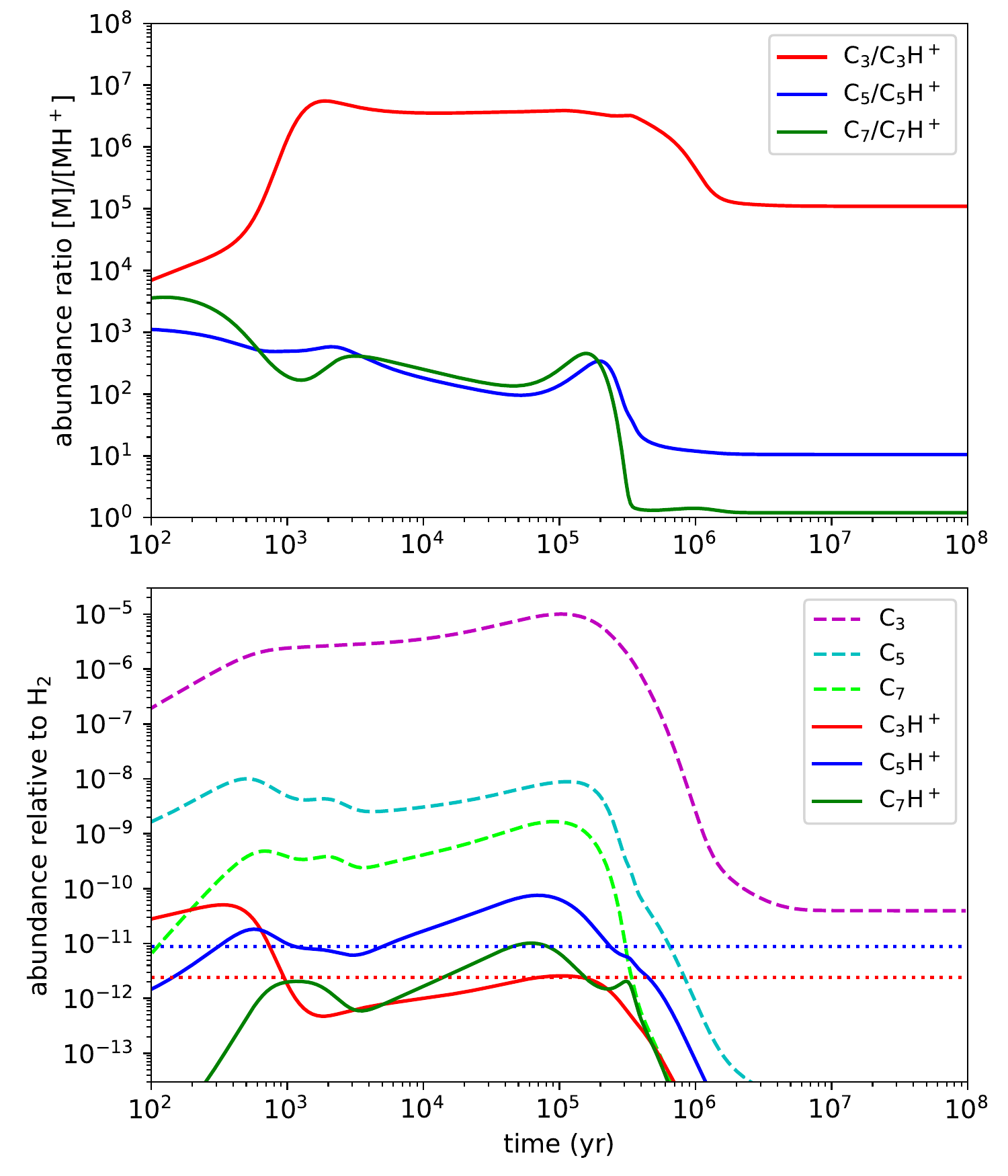}
\caption{Chemical models for the C$_n$H$^+$ species. ({\it upper}) Computed abundance ratio between the clusters C$_3$, C$_5$ and C$_7$ and their protonated
forms C$_n$H$^+$ as a function of time. ({\it lower}) Abundances relative to H$_2$ for
the carbon clusters and their protonated derivatives. The dotted horizontal lines correspond to the abundances observed 
in \mbox{TMC-1} for C$_3$H$^+$ and C$_5$H$^+$.}
\label{fig:abun}
\end{figure}

To investigate the chemistry behind the formation of C$_3$H$^+$ and C$_5$H$^+$ , we built a pseudo-time-dependent 
gas-phase chemical model of a cold dark cloud using a chemical network largely based on the RATE12 network from the 
UMIST database \citep{McElroy2013}. The model is similar to the models presented by \cite{Agundez2015}, \cite{Marcelino2020}, 
and \cite{Cabezas2022}. The results from the chemical model are shown in Fig.~\ref{fig:abun}, where one should focus on 
the so-called early time (10$^5$-10$^6$ yr), at which calculated and observed abundances in \mbox{TMC-1} show the best 
overall agreement (see, e.g., \citealt{Agundez2013}). The pure carbon clusters with an odd number of carbon atoms C$_3$, 
C$_5$, and C$_7$ are calculated to be abundant. In particular, C$_3$ is predicted to be very abundant, with a peak 
abundance as high as $\sim$\,10$^{-5}$ relative to H$_2$. The carbon clusters C$_5$ and C$_7$ have calculated peak 
abundances in the range 10$^{-9}$-10$^{-8}$ relative to H$_2$. For these neutral carbon clusters, the chemical model 
indicates that the larger the size, the lower the abundance. In the case of the ions C$_3$H$^+$, C$_5$H$^+$, and C$_7$H$^+$, 
which can be seen as the protonated forms of C$_3$, C$_5$, and C$_7$, respectively, their calculated early-time abundances 
do not show the trend of decreasing abundance with increasing size. In fact, the smallest member, C$_3$H$^+$, has the 
lowest calculated abundance, while the medium-sized ion C$_5$H$^+$ shows the highest abundance. This is in agreement with 
the observations of \mbox{TMC-1}, which indicate that C$_5$H$^+$ is more abundant than C$_3$H$^+$. The underlying reason 
of this behavior is the different reactivity of C$_3$H$^+$ and C$_5$H$^+$ with H$_2$.

In a simplified chemical scheme, protonated molecules are mostly formed by proton transfer to the corresponding neutral 
counterpart, while they are mainly destroyed through dissociative recombination with electrons (e.g., \citealt{Agundez2015}). 
However, in the case of the protonated forms of the carbon clusters C$_3$, C$_5$, and C$_7$, the behavior is somewhat different. 
Moreover, the chemical processes of formation and destruction are very different for C$_3$H$^+$ and for the larger analogs 
C$_5$H$^+$ and C$_7$H$^+$. The underlying reasons are that the abundance of C$_3$ is far higher than that of C$_5$ and 
C$_7$, and the reactivity of C$_3$H$^+$ with H$_2$ is far stronger than that of C$_5$H$^+$ and C$_7$H$^+$. According to the 
chemical model, C$_3$H$^+$ is mainly formed through proton transfer from HCO$^+$ and H$_3$O$^+$ to C$_3$, following the usual 
pathway of many other protonated molecules. However, the main destruction process of C$_3$H$^+$ does not involve the dissociative 
recombination with electrons, but the reaction with H$_2$. This reaction occurs fast through radiative association, leading to 
the linear and cyclic isomers of the ion C$_3$H$_3^+$, which are key intermediates in the synthesis of C$_3$H$_2$ isomers 
(see, e.g., \citealt{Loison2017}). In the cases of C$_5$H$^+$ and C$_7$H$^+$, the main formation pathways do not involve 
proton transfer to C$_5$ and C$_7$, but other ion-neutral reactions such as C$_5^+$ + H$_2$, C$^+$ + C$_4$H$_2$, and C + 
C$_4$H$_2^+$ in the case of C$_5$H$^+$, and C$_7^+$ + H$_2$, C$^+$ + C$_6$H$_2$, and C + C$_6$H$_2^+$ for C$_7$H$^+$. In the case 
of C$_3$H$^+$, the high abundance of C$_3$ makes the route involving proton transfer very efficient, while the lower 
abundances of C$_5$ and C$_7$ make the proton transfer pathways less efficient than other ion-neutral routes. The 
destruction of C$_5$H$^+$ and C$_7$H$^+$ is different from that of C$_3$H$^+$ because the former shows a much lower reactivity 
with H$_2$ than the latter. As a consequence, C$_5$H$^+$ and C$_7$H$^+$ are mostly destroyed through dissociative 
recombination with electrons and reaction with neutral atoms, as occurs for many other protonated molecules. The 
different chemistry of C$_3$H$^+$, compared to that of C$_5$H$^+$ and C$_7$H$^+$ directly translates into the calculated 
abundances and neutral-to-protonated abundance ratios. While the calculated C$_5$/C$_5$H$^+$ and C$_7$/C$_7$H$^+$ ratios 
are in the range 1-10$^3$ found for other protonated molecules (e.g., \citealt{Agundez2015,Cernicharo2020a,Cabezas2022}), 
the calculated C$_3$/C$_3$H$^+$ is much higher (see the top panel in Fig.~\ref{fig:abun}). Moreover, the calculated 
abundance of C$_3$H$^+$ remains below that of C$_5$H$^+$. It is worth noting, however, that in spite of the reactivity of 
C$_3$H$^+$ with H$_2$, its calculated abundance does not drop to negligible levels because the abundance of its 
main precursor, the carbon cluster C$_3$, is very high.

Last, if we trust the chemical model, then it is predicted that C$_7$H$^+$ should be present in \mbox{TMC-1 as well,} with 
an abundance a few times below that of C$_5$H$^+$. The current sensitivity of the QUIJOTE line survey, however, does not allow us
to identify C$_7$H$^+$, but the improvement in the sensitivity planned for the coming years could allow us to detect this species and 
confirm the above chemical scenario.

\section{Conclusions}

We reported the discovery of the cation C$_5$H$^+$ in TMC-1. We also 
reported the first detection of the
cation C$_3$H$^+$ in a cold starless core. It was previously only observed in the direction
of PDRs and diffuse interstellar clouds. 
The assignment of the four observed, harmonically related lines, 
to C$_5$H$^+$ was based on accurate ab initio
calculations, which permit us to rule out C$_5$H$^-$ as the possible
carrier of these lines. Our chemical model reproduces the observations satisfactorily and explains the 
lower abundance of C$_3$H$^+$ with respect to C$_5$H$^+$ as due to the much higher reactivity of the 
former with respect to the latter. C$_7$H$^+$ is
predicted to have an abundance a few times below that of C$_5$H$^+$. It might therefore be detectable with future QUIJOTE data with a higher signal-to-noise ratio.

\begin{acknowledgements}
We thank ERC for funding
through grant ERC-2013-Syg-610256-NANOCOSMOS. M.A. thanks MICIU for grant 
RyC-2014-16277. We also thank Ministerio de Ciencia e Innovaci\'on of Spain (MICIU) for funding support through projects
PID2019-106110GB-I00, PID2019-107115GB-C21 / AEI / 10.13039/501100011033, 
and PID2019-106235GB-I00. 

\end{acknowledgements}

%\begin{appendix}

%\onecolumn
%\end{appendix}
\end{document}